# ERROR GRADIENT-BASED VARIABLE-LP NORM CONSTRAINT LMS ALGORITHM FOR SPARSE SYSTEM IDENTIFICATION

*Yong Feng[1,2], Fei Chen[2], Rui Zeng[1], Jiasong Wu[1], Huazhong Shu[1]*
[1] School of CSE, Southeast University, Nanjing, 210096, China
[2] Department of EEE, South University of Science and Technology of China, Shenzhen, 518055, China
{fengy, fchen}@sustc.edu.cn, jswu@seu.edu.cn

**ABSTRACT**

Sparse adaptive filtering has gained much attention due to its wide applicability in the field of signal processing. Among the main algorithm families, sparse norm constraint adaptive filters develop rapidly in recent years. However, when applied for system identification, most priori work in sparse norm constraint adaptive filtering suffers from the difficulty of adaptability to the sparsity of the systems to be identified. To address this problem, we propose a novel variable *p*-norm constraint least mean square (LMS) algorithm, which serves as a variant of the conventional $L_p$-LMS algorithm established for sparse system identification. The parameter *p* is iteratively adjusted by the gradient descent method applied to the instantaneous square error. Numerical simulations show that this new approach achieves better performance than the traditional $L_p$-LMS and LMS algorithms in terms of steady-state error and convergence rate.

## 1. INTRODUCTION

Recent years have witnessed a rush of interest in sparse adaptive filtering, e.g., sparse system identification [1] and sparse channel estimation [2], which is mainly motivated by the research of the least absolute shrinkage and selection operator (LASSO) [3] and compressive sensing (CS) [4]. The well-known least mean square (LMS) algorithm [5] has been widely used in adaptive filtering due to its computational simplicity, and the family of norm constraint LMS algorithms which have attracted much attention currently, has exhibited higher performance than the standard LMS and greater robustness against additive noise in estimating sparse systems [6-9].

To improve the performance of the standard LMS algorithm, researchers have proposed many norm constraint LMS algorithms, for instance, $L_1$-norm penalty LMS ($L_1$-LMS) [6, 10], $L_0$-norm penalty LMS ($L_0$-LMS) [7, 11, 12], $L_p$ norm penalty LMS ($L_p$-LMS) [8] and $L_p$ norm like LMS ($L_{pl}$-LMS) [9], where different corresponding norm constraints are integrated into the cost function of the conventional LMS algorithm, respectively, to increase the convergence speed and/or decrease the mean square error (MSE). However, these existing algorithms generally suffer from the difficulty of adaptability to the sparsity of the systems to be identified, due to the lack of any adjustable factors [9].

In this paper, we develop a new *p* norm constraint LMS algorithm with a variable *p* to address the above-mentioned challenge. It is achieved by iteratively adjusting *p* along the negative gradient direction of the instantaneous square error (SE) with respect to *p*, which leads to an optimal *p*-norm constraint for the $L_p$-LMS. Numerical simulation results show that the proposed algorithm has better performance than the standard LMS and $L_p$-LMS algorithms.

The organization of this paper is as follows: an overview of the standard LMS and norm constraint LMS algorithms is presented in Section II. The proposed variable *p*-norm constraint LMS algorithm is then detailed in Section III. The numerical validation on simulated scenarios is given in Section IV in the setting of sparse system identification. Finally Section V concludes the paper.

## 2. EXISTING LMS ALGORITHMS

Throughout this paper, let $y_k$ be the output of a finite impulse response (FIR) system with an additive noise $n_k$ at time *k*, which can be written as follows:

$$y_k = \mathbf{w}^T \mathbf{x}_k + n_k \qquad (1)$$

where the weight vector **w** of length *N* is the sparse impulse response of the unknown system, and $(\cdot)^T$ denotes the transpose operator. $\mathbf{x}_k$ represents the stationary input vector with zero mean and covariance matrix **R**, consisting of the last *N* input signal samples, i.e., $\mathbf{x}_k = [x_k, x_{k-1}, \cdots, x_{k-N+1}]^T$. $n_k$ is a stationary noise process with zero mean and variance $\sigma_k^2$. Given the input $\mathbf{x}_k$ and output $y_k$ following the above linear system model, the problem is to estimate the weight vector **w**.

This work was supported by the National Basic Research Program of China under Grant 2011CB707904, by the NSFC under Grants 61571213, 61201344 and 61271312, and by NSF of Jiangsu Province under Grant BK2012329.

In the standard LMS [5], the cost function $J_k$ to be minimized is defined as:

$$J_k \triangleq e_k^2 / 2 \quad (2)$$

where $e_k$ denotes the instantaneous error between the output and desired response, i.e., $e_k = y_k - \mathbf{w}_k^T \mathbf{x}_k$, and $\mathbf{w}_k = [w_{k,1}, w_{k,2}, ..., w_{k,N}]^T$ is the estimated weight vector of the filter at time $k$. Note that the "1/2" here is taken just for the convenience of notation. Then the update equation is written as:

$$\mathbf{w}_{k+1} = \mathbf{w}_k - \mu \frac{\partial J_k}{\partial \mathbf{w}_k} = \mathbf{w}_k + \mu e_k \mathbf{x}_k \quad (3)$$

where $\mu$ is the step size such that $0 < \mu < \lambda_{max}^{-1}$ with $\lambda_{max}$ being the maximum eigenvalue of $\mathbf{R}$. The gradient descent method employed here assures the convergence to the optimum under the aforementioned condition on $\mu$, due to the convexity of the cost function.

Considering a common scenario that the weight vector $\mathbf{w}$ is sparse, i.e., most of its coefficients are exactly or nearly zeros, several sparsity-aware modifications of the LMS algorithm have been developed to improve the performance by exploiting the sparse prior information. For example, $L_1$-LMS, $L_0$-LMS and $L_p$-LMS introduce the $L_1$, $L_0$ and $L_p$ norms of the weight vector into the cost function of the standard LMS, respectively. For $L_p$-LMS [8], the new cost function becomes

$$J_{k,p} \triangleq e_k^2 / 2 + \gamma_p \|\mathbf{w}_k\|_p \quad (4)$$

where the $L_p$ norm is defined as $\|\mathbf{w}_k\|_p \triangleq \left( \sum_{i=1}^{N} |w_{k,i}|^p \right)^{1/p}$ with $0 < p < 1$, and $\gamma_p$ is a constant weight assigned to the penalty term and is determined by the trade-off between the convergence speed and estimation error. Consequently, its update equation is derived as:

$$\mathbf{w}_{k+1} = \mathbf{w}_k + \mu e_k \mathbf{x}_k - \rho_p \frac{\|\mathbf{w}_k\|_p^{1-p} \operatorname{sgn}(\mathbf{w}_k)}{\varepsilon_p + |\mathbf{w}_k|^{1-p}} \quad (5)$$

where $\rho_p = \mu \gamma_p$, $\varepsilon_p$ is a constant which is imposed to bound the last term in the situation when an entry of $\mathbf{w}_k$ approaches zero and sgn(x) is the sign function, which is zero for x = 0, 1 for x > 0 and -1 for x < 0 while sgn($\mathbf{w}_k$) applies to each element of $\mathbf{w}_k$, respectively. Note that the cost function of the $L_p$-LMS is not convex such that the convergence and consistency analysis is problematic [13].

More recently, a new sparse LMS algorithm was proposed by Wu and Tong [9], where the $p$-norm like constraint [14] is integrated into the cost function of the LMS to exert a zero attractor to the weight updating equation as follows:

$$\mathbf{w}_{k+1} = \mathbf{w}_k + \mu e_k \mathbf{x}_k - \rho_{pl} \frac{p \operatorname{sgn}(\mathbf{w}_k)}{\varepsilon_{pl} + |\mathbf{w}_k|^{1-p}} \quad (6)$$

where constants $\rho_{pl}$ and $\varepsilon_{pl}$ act the same as those in the $L_p$-LMS, respectively.

## 3. PROPOSED ALGORITHMS

The $L_p$-LMS can achieve better performance than the standard LMS, comparable to or sometimes better than the $L_1$-LMS, $L_0$-LMS and $L_{pl}$-LMS [8]. However, it is characterized by the difficulty of adaptability to the sparsity of the systems, due to the lack of any adjustable factors. Therefore this study developed a new $L_p$-LMS algorithm with a variable $p$ to overcome the above limitation, termed as $L_{vp(GSE)}$-LMS, which is achieved by iteratively adjusting $p$ by the gradient descent method of the instantaneous SE with respect to $p$, thus leading to an optimal $p$-norm constraint for the $L_p$-LMS. The update equation of variable $p$ for $L_{vp(GSE)}$-LMS is:

$$p_{k+1} = p_k - \delta \operatorname{GSE}_{p,k} \quad (7)$$

where $\delta$ is a constant factor to control the step size of descent gradient method, which plays the same role as $\mu$ in the weight update of the LMS algorithm, and $\operatorname{GSE}_{p,k}$ is the gradient of the SE function $\operatorname{SE}_k \triangleq e_k^2$ with respect to $p$, which can be derived as:

$$\operatorname{GSE}_{p,k} = \frac{\partial \operatorname{SE}_k}{\partial p} = \|\mathbf{w}_{k-1}\|_p^{1-p} \mathbf{A}(p) \{ \mathbf{B}(p) [C(p)\mathbf{D}(p) + \mathbf{E}(p)] \} \quad (8)$$

where

$$\mathbf{A}(p) = 2\rho e_k \mathbf{x}_k^T, \mathbf{B}(p) = \frac{\operatorname{sgn}(\mathbf{w}_{k-1})}{\left( \varepsilon + |\mathbf{w}_{k-1}|^{1-p} \right)^2},$$

$$C(p) = \frac{(1-p) \sum \left( |[\mathbf{w}_{k-1}]_i|^p \ln |[\mathbf{w}_{k-1}]_i| \right)}{\sum |[\mathbf{w}_{k-1}]_i|^p} - \ln \|\mathbf{w}_{k-1}\|_p,$$

$$\mathbf{D}(p) = (\varepsilon + |\mathbf{w}_{k-1}|^{1-p})/p, \mathbf{E}(p) = |\mathbf{w}_{k-1}|^{1-p} \ln |\mathbf{w}_{k-1}|.$$

where $|\mathbf{w}_{k-1}|$ applies the absolute value to each component and $[\mathbf{w}_{k-1}]_i$ represents the $i$th element of $\mathbf{w}_{k-1}$, $i = 1,2,...,N$.

Unfortunately, we cannot guarantee convexity of the SE function with respect to $p$. Inspired by Wu and Tong's work [9], this study takes the sign and smoothed version of the gradient to impose on the gradient descent derivation to avoid this problem as much as possible, i.e.,

$$p_{k+1} = p_k - \delta_k \operatorname{sgn}\left( \frac{1}{T} \sum_{i=k-T+1}^{k} \operatorname{GSE}_{p,i} \right) \quad (9)$$

Additionally, though δ is determined by the trade-off between adaptation speed and optimization accuracy, we notice that the proposed $L_{vp(GSE)}$-LMS algorithm converges fast as well as keeps stable if the parameter $δ$ varies from a larger initial value to a smaller stable value during the iterations. Hence, a simple scheme of a variable $δ$ is given by:

$$\delta_{k+1} = \delta_k - u \quad (10)$$

where $u$ is a very small step size. To summarize the proposed algorithm, the pseudo-codes for Matlab are listed in Table 1.

Table 1. Pseudo-codes of the $L_{vp(GSE)}$-LMS algorithm.

| Given | $\mu, \rho, N, T, \varepsilon, w, x, n, L, u$ |
|---|---|
| Initial | $\mathbf{w}_0 = \text{zeros}(N,1), p_0, \mathbf{Y} = \text{zeros}(1,L), \delta_0$ |
| For | $i = 1, 2, ..., (L-N+1)$ |
| | $y_k = \mathbf{w}^T \mathbf{x}_k + n_k$ |
| | $e_k = y_k - \mathbf{w}_k^T \mathbf{x}_k$ |
| | $\mathbf{w}_{k+1} = \mathbf{w}_k + \mu e_k \mathbf{x}_k - \rho \dfrac{\|\mathbf{w}_k\|_p^{1-p} \operatorname{sgn} \mathbf{w}_k}{\varepsilon + |\mathbf{w}_k|^{1-p}}$ |
| | $p_{k+1} = p_k - \delta_k \operatorname{sgn}\{2\rho \|\mathbf{w}_k\|_p^{1-p} e_{k+1} \mathbf{x}_{k+1}^T \ast \{\ldots$ |
| | $\operatorname{sgn} \mathbf{w}_k ./ (\varepsilon + |\mathbf{w}_k|^{1-p})^2 .\ast [\ldots$ |
| | $(1-p)\sum (|[\mathbf{w}_k]_i|^p \ln |[\mathbf{w}_k]_i|) / \ldots$ |
| | $\sum |[\mathbf{w}_k]_i|^p - \ln \|\mathbf{w}_k\|_p \ldots$ |
| | $\ast (\varepsilon + |\mathbf{w}_k|^{1-p}) / p \ldots$ |
| | $+ |\mathbf{w}_k|^{1-p} .\ast \ln |\mathbf{w}_k| \ldots$ |
| | $]\ldots$ |
| | $\}$; |
| end | $\delta_{k+1} = \delta_k - u$; |

Table 2. Parameters of the algorithms in the simulations.

| Algorithms | $\mu$ | $\rho$ | $\varepsilon$ | $p$ | $\delta$ | $T$ |
|---|---|---|---|---|---|---|
| LMS | | / | / | / | / | / |
| $L_p$-LMS | | | | 0.5 | / | / |
| $L_{vp(GSE)}$-LMS | $5 \times 10^{-2}$ | $5 \times 10^{-4}$ | $5 \times 10^{-2}$ | $p_0 = 1$ * | 0.01** | 5 |
| $L_{vp(GSD)}$-LMS | | | | $p_0 = 0.5$ | 0.05*** | |

\* $p_0$ is the initial value of variable $p$.
\*\* $\delta = 0.01, 0.005, 0.003, 0.001$ and 0 for the 1st, 2nd, 3rd, 4th and last 100 iterations, respectively.
\*\*\* $\delta = 0.05, 0.03, 0.02, 0.01$ and 0 for the 11-30th, 31-50th, 51-70th and 71-90th iterations while 0.005 for 90-200th and 0.001 for the rest, respectively.

Similarly, we might also develop a new $L_p$-norm like LMS with a variable $p$ with the corresponding gradient $\text{GSE}_{pl,k}$ of the SE function $\text{SE}_{k,pl}$ with respect to $p$, called $L_{vpl(GSE)}$-LMS, whose core $p$-varying equation is represented as:

$$\text{GSE}_{pl,k} = \frac{\partial \text{SE}_{k,pl}}{\partial p} = p\mathbf{A}(p)\{\mathbf{B}(p)[\mathbf{D}(p) + \mathbf{E}(p)]\} \quad (11)$$

which is much simpler for computation than that in the $L_{vp(GSE)}$-LMS algorithm.

Moreover, we also derive another modification of $L_p$-LMS algorithm with variable $p$ based on the gradient of the instantaneous square deviation (SD) called $L_{vp(GSD)}$-LMS as comparison, where the gradient of the SD function $\text{SD}_k \triangleq \|\mathbf{w} - \mathbf{w}_k\|_2^2$ with respect to $p$ is derived as:

$$\text{GSD}_{p,k} = \frac{\partial \text{SD}_k}{\partial p}$$
$$= 2\rho_p \|\mathbf{w}_{k-1}\|_p^{1-p} \sum_{i=1}^N \{([\mathbf{w}]_i - [\mathbf{w}_k]_i)\mathbf{B}(p)[C(p)\mathbf{D}(p) + \mathbf{E}(p)]\} \quad (12)$$

Note that as the real weight vector $\mathbf{w}$ is involved in the above equation, this method is never able to be employed in the system identification problems. However, it can be utilized to set up a range pole for choosing the optimal (or sub-optimal) value of $p$ in this paper, and might be helpful in adjusting parameters in practical sparse LMS problems. As shown in the simulations later, this approach will always converge to a different but fixed $p$ and achieve the best performance for all sparsity levels, which is obviously due to the exploitation of the real weight vector $\mathbf{w}$ that is actually to be identified.

Furthermore, as the computation complexity of the decimal exponential $|[\mathbf{w}_k]_i|^{1-p_k}$ is such high that the Newton iteration idea can be exploited to simplify the gradient descent optimization in both the weight updates and the parameter $p$ iterations [15]. Let $g \triangleq |[\mathbf{w}_k]_i|^{1-p_k}$, then the iteration can be written as:

$$g_{j+1} = g_j - \frac{g_j^{1/\delta} - |[\mathbf{w}_k]_i|^{(1-p_k)/\delta}}{g_j^{1/\delta-1}/\delta} \quad (13)$$

where $j$ usually only takes a small number like 3 or 4 times to meet the general performance requirement sufficiently.

For the analysis of convergence, we define the misalignment vector $\mathbf{v}_k = \mathbf{w}_k - \mathbf{w}$. Substituting the misalignment vector into Eqs. (5) and (6) and taking expectations on both sides of the equations, then we obtain

$$\mathrm{E}[\mathbf{v}_{k+1}]_p = (\mathbf{I} - \mu\mathbf{R})\mathrm{E}[\mathbf{v}_k]_p - \rho_p \frac{\|\mathbf{w}_k\|_p^{1-p} \operatorname{sgn}(\mathbf{w}_k)}{\varepsilon_p + |\mathbf{w}_k|^{1-p}} \quad (14)$$

for the $L_{vp}$-LMS and

$$\mathrm{E}[\mathbf{v}_{k+1}]_{pl} = (\mathbf{I} - \mu\mathbf{R})\mathrm{E}[\mathbf{v}_k]_{pl} - \rho_{pl} \frac{p \operatorname{sgn}(\mathbf{w}_k)}{\varepsilon_p + |\mathbf{w}_k|^{1-p}} \quad (15)$$

for the $L_{vpl}$-LMS, respectively, where $\mathbf{I}$ is the identity matrix and $\rho_{pl} p \operatorname{sgn}(\mathbf{w}_k) / \varepsilon_p + |\mathbf{w}_k|^{1-p}$ is bounded between $\pm \rho_{pl} p / \varepsilon_p + |\mathbf{w}_k|^{1-p}$. Therefore, $\mathrm{E}[v_k]_{pl}$ for $L_{vpl}$-LMS converges if $0 < \mu < \lambda_{\max}^{-1}$, whereas the convergence of $\mathrm{E}[v_k]_p$ for $L_{vp}$-LMS remains problematic, due to the nonconvexity of its cost function as mentioned above.

## 4. SIMULATIONS

Numerical simulations are carried out to test the performances of the proposed $L_{vp(GSE)}$-LMS algorithm in terms of the steady-state mean square deviation (MSD, defined as $\text{MSD}_k = \mathrm{E}[\|\mathbf{w} - \mathbf{w}_k\|_2^2]$) and convergence rate, and the results are compared with those from standard LMS, the $L_p$-LMS and $L_{vp(GSD)}$-LMS algorithms in sparse system identification with different sparsity levels.

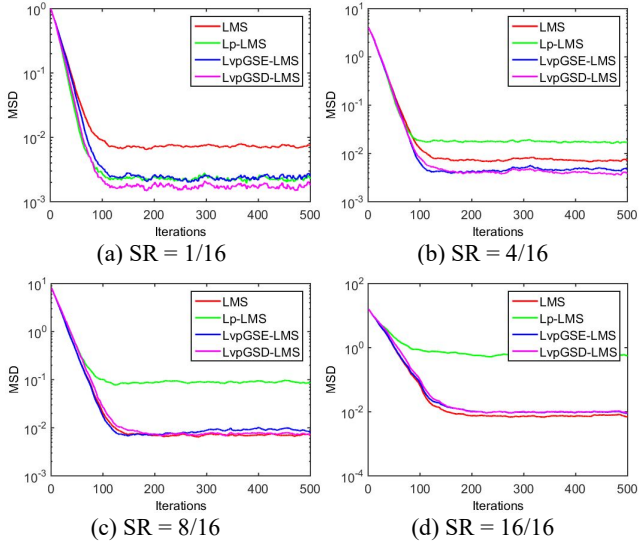

Fig. 1. MSD curves of different algorithms with different sparsity levels. The SNR level was set to 20 dB.

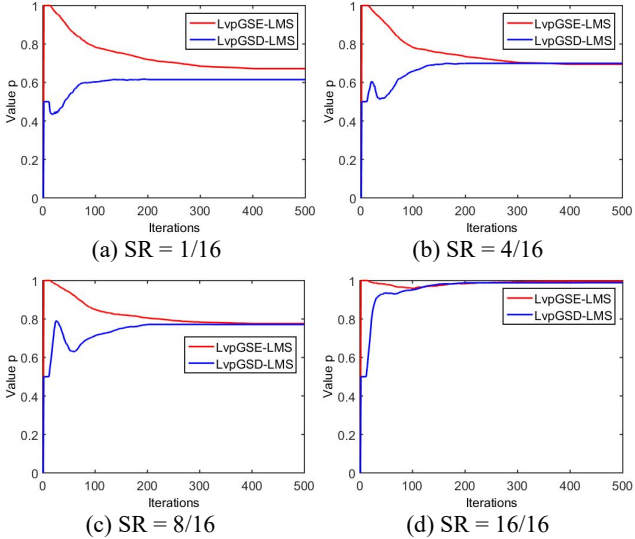

Fig. 2. The iterations of $p$ with different sparsity levels. Corresponding the the cases in Fig.1, respectively.

We identify a 16-tap sparse unknown system with 1, 4, 8, or 16 taps assumed to be nonzero. That is, the sparsity ratio (SR) is set to be 1/16, 4/16, 8/16 and 16/16, respectively. The positions of nonzero taps are chosen randomly and the values are chosen from a zero-mean Gaussian distribution with unit variance. The input signal and observed noise are both assumed to be white Gaussian processes of length 500 with zero mean and variances 1 and 0.01, respectively, i.e., the signal noise ratio (SNR) is 20 dB. Other parameters are carefully selected as listed in Table 2. All the simulations are obtained by 200 Monte-Carlo runs.

Figure 1 shows the MSD curves of these algorithms versus the number of iterations employed to identify the unknown system with different sparsity levels of SR = 1/16, 4/16, 8/16 and 16/16. As we can see from Fig. 1, the $L_{vp\text{(GSE)}}$-LMS, $L_{vp\text{(GSD)}}$-LMS and $L_p$-LMS all generally yield faster convergence rate than the standard LMS does when the system is very sparse, i.e., SR=1/16, due to the sparsity exploitation stated previously. However, they have different performance enhancements or deteriorations compared to the standard LMS while the sparsity of the system varies, which are determined by the related penalty control factors like $\gamma$, $p$ as well as SNR of the input. Overall, the $L_{vp\text{(GSE)}}$-LMS performs better than the $L_p$-LMS for all the different sparsity and very close to the $L_{vp\text{(GSD)}}$-LMS algorithm, though it does not work well under the semi-sparse and non-sparse condition as expected, and neither does the $L_p$-LMS, which performs very bad in these cases. Accordingly, the performance of the negative gradient $p$ parameter optimization is presented in Fig. 2, showing how $p$ iteratively converges from the initial value $p_0 = 1$. As we can see from Fig. 2, the varying value of $p$ almost always converges very close to that of the $L_{vp\text{(GSD)}}$-LMS for all sparsity ratios, which demonstrates our proposed $L_{vp\text{(GSE)}}$-LMS algorithm yields a better $p$ for $L_p$-LMS. In addition, the $p$ value is very close to the optimal (or sub-optimal) one in $L_{vp\text{(GSD)}}$-LMS, at the cost of a little higher but still acceptable computation complexity.

## 5. CONCLUSION

In order to exploit the inferior sparse information effectively to improve the performance of the system identification, and conquer the problem that most priori proposed algorithms in the area of sparse norm constraint adaptive filtering suffer from the difficulty of adaptability to the sparsity of system, the present work in this paper develops a novel $L_p$ norm constraint LMS algorithm coined as the $L_{vp\text{(GSE)}}$-LMS, in which the variable $p$ is iteratively adapted to the gradient descent of the instantaneous square error function. Similarly, a new $L_p$-LMS algorithm with variable $p$ based on the gradient of square deviation, called the $L_{vp\text{(GSD)}}$-LMS, is also derived as the range pole and companion in the estimation of parameter $p$. Numerical simulation results show that the proposed $L_{vp\text{(GSE)}}$-LMS algorithm achieves better performance than the traditional $L_p$ norm constraint LMS and the standard LMS algorithm, since it demonstrates a better sparsity exploration and also tolerates better on different sparse levels.

Our future work will focus more systematically on the parameters optimization of $L_p$-LMS and other similar algorithms in sparse system identification settings. We will explore more details in the relationship among the sparsity level of system, the weight of sparse norm constraint, the signal-noise ratio of the input, etc., to develop new algorithms that adapt to the sparsity ratio better. Furthermore, the method employed in this work can be further extended to other $L_p$-norm and $L_p$-norm-like related adaptive filters as well.